\documentclass[conference,10pt]{IEEEtran}
\usepackage[utf8]{inputenc}
\def\BibTeX{{\rm B\kern-.05em{\sc i\kern-.025em b}\kern-.08em
    T\kern-.1667em\lower.7ex\hbox{E}\kern-.125emX}}

\usepackage{graphicx}
\usepackage{color}
\usepackage{cite}
\usepackage{amsmath}
\usepackage[caption=false]{subfig}
\usepackage{amssymb}
\usepackage{comment}
\usepackage{mathtools}
\usepackage{tabulary}
\usepackage{acronym}
\usepackage{upgreek}
\usepackage{algorithm}
\usepackage{algpseudocode}
\usepackage{placeins}
\usepackage[super]{nth}
\usepackage{dsfont,flushend}
\usepackage{hyperref}

\title{Autoregressive Attention Neural Networks for Non-Line-of-Sight User Tracking with \\
      Dynamic Metasurface Antennas}

\author{\IEEEauthorblockN{Kyriakos Stylianopoulos$^1$, Murat Bayraktar$^2$, Nuria Gonz\'{a}lez-Prelcic$^2$, and George C. Alexandropoulos$^1$}\\
\IEEEauthorblockA{$^1$Department of Informatics and Telecommunications,
National and Kapodistrian University of Athens\\
Panepistimiopolis Ilissia, 15784 Athens, Greece\\
$^2$Department of Electrical and Computer Engineering, North Carolina State University, 27695 Raleigh, USA\\
e-mails: \{kstylianop, alexandg\}@di.uoa.gr,  \{mbayrak, ngprelcic\}@ncsu.edu \\
}}

\begin{document}

\maketitle

\begin{abstract}
User localization and tracking in the upcoming generation of wireless networks have the potential to be revolutionized by technologies such as the Dynamic Metasurface Antennas (DMAs). Commonly proposed algorithmic approaches rely on assumptions about relatively dominant Line-of-Sight (LoS) paths, or require pilot transmission sequences whose length is comparable to the number of DMA elements, thus, leading to limited effectiveness and considerable measurement overheads in blocked LoS and dynamic multipath environments. In this paper, we present a two-stage machine-learning-based approach for user tracking, specifically designed for non-LoS multipath settings.
A newly proposed attention-based Neural Network (NN) is first trained to map noisy channel responses to potential user positions, regardless of user mobility patterns. This architecture constitutes a modification of the prominent vision transformer, specifically modified for extracting information from high-dimensional frequency response signals. As a second stage, the NN's predictions for the past user positions are passed through a learnable autoregressive model to exploit the time-correlated channel information and obtain the final position predictions.
The channel estimation procedure leverages a DMA receive architecture with partially-connected radio frequency chains, which results to reduced numbers of pilots. The numerical evaluation over an outdoor ray-tracing scenario illustrates that despite LoS blockage, this methodology is capable of achieving high position accuracy across various multipath settings.

\end{abstract}

\section{Introduction}

\let\thefootnote\relax\footnotetext{This work has been supported by the EU H2020 RISE-6G project under grant number 10101701.}

Metasurface-based antennas, such as Reconfigurable Intelligent Surfaces (RISs) \cite{RIS_Overview} \cite{huang2019reconfigurable} and Dynamic Metasurface Antenna arrays (DMAs) \cite{DMA_for_MIMO,FD_HMIMO_2023} are key enablers of smart radio environments \cite{RISE6G_COMMAG}, and can offer benefits across various network objectives, such as localization, sensing, and Radio Frequency (RF) mapping. In many recent cases, RISs have been endowed with receive RF Chains (RFCs) and play the role of large-aperture receivers, or sensors, of uplink signals, either by deploying multiple single-RF metasurfaces~\cite{MultiRIS1RFC} or partially-connected multi-RFC architectures with reflecting~\cite{partially_connected_receiving_RIS} and hybrid reflecting/sensing~\cite{HRIS_Mag} meta-elements.

When solely reflecting surfaces are considered in bistatic sensing architectures \cite{LIS_Positioning, RIS_reflection_modulation,CS_3D_Distributed, RIS_sidelink}, a two-stage procedure is most commonly adopted in estimating channel or environment components (e.g., Angles of Arrivals (AoAs)), and then, performing the position estimation.
Typically, the information extraction over the signals is carried out via subspace based algorithms (e.g., MUSIC \cite{MUSIC} and its extensions \cite{RootMUSIC, DA-MUSIC}) or Compressed Sensing (CS) approaches (e.g. \cite{CS_3D_Distributed, LIS_Positioning}), while the position estimation can be obtained via minimizing the Cramér-Rao Bound (CRB) \cite{CRB_DoA_Review}.
Recent methodologies stemming from Machine Learning (ML) are either designed to augment conventional approaches, such as the Deep-Augmented MUSIC approach \cite{DA-MUSIC}, or may focus on active sensing setups where the algorithms try to find favorable metasurface configurations that lead to sufficient estimations.
To that end, very recently, a Long-Short-Term Memory (LSTM) neural network was proposed in \cite{ActiveSensingICC23} for beamforming toward the static user position, and a similar strategy based on reinforcement learning was used in \cite{MetaSensingDRL} to localize passive objects.

While subspace/CS-based and ML approaches have demonstrated great potential, they often rely on extensive signal sampling to perform estimations (i.e., in the order of the number of RIS elements \cite{Samarakoon_2020,DRL_RIS_2020,MultiRIS1RFC,RIS_sidelink,Alexandropoulos2022Pervasive}) that may well exceed the coherence time of the channel.
This problem is only exacerbated by the fact that, switching configurations in metasurfaces introduce non-negligible delays (which may be even comparable to the channel coherence frame~\cite{ETSI_RIS_1}), and that most of the available techniques have non-linear computational complexity~\cite{RIS_Overview}. Besides, the vast majority of localization approaches, even the ones that follow similar ML-based methodologies \cite{Nuria_Sparse_Recovery, Nuria_ChanFormer}, assume Line-of-Sight (LoS) propagation environments between the receiver/metasurface and the user \cite{Nuria_Sparse_Recovery}, or may entail large sensing and preprocessing overheads \cite{Nuria_ChanFormer, Gante_DL_2020}.

\textbf{Contributions:} Taking the above considerations into account, in this paper, we design a monostatic sensing system for user tracking in multipath non-LoS environments, where a DMA with a limited number of RFCs observes an equal number of pilot signals to perform a noisy estimation of the channel.
A novel modified Multi-Head Self-Attention (mMHSA) Neural Network (NN) architecture is proposed that learns a mapping between the channel vectors and the (static) user positions within the considered environment. To perform tracking, we furthermore train an AutoRegressive (AR) model that receives as input the time-agnostic predictions of the mMHSA network along sampled user trajectories, and outputs a smoothed version of the predictions that better fit the mobility patterns of the users. 

\textbf{Notation}: Vectors (matrices) are given in bold lowercase (bold uppercase) typeface and calligraphic letters denote sets. $(\cdot)^T$ and $(\cdot)^H$ return matrix transposition and hermitian transposition, respectively, while $\|\cdot\|$ is the Euclidean norm and $|\cdot|$ provides a set's cardinality. $\mathcal{CN}(\cdot,\cdot)$ denotes the complex normal distribution and $\jmath\triangleq \sqrt{-1}$. For a vector $\boldsymbol{v}$, $\boldsymbol{v}_{i:j}$ denotes the sliced-vector constructed by taking the elements of $\boldsymbol{v}$ from the $i$-th position up to, and including, the $j$-th position.

\section{System Model and Objective}
\subsection{System Model}
We consider the uplink of a communication system, in which a single-antenna user moves along pseudorandom 2D trajectories and periodically transmits localization requests to a fixed-position planar DMA-based access point comprising $N$ meta-elements, each with a tunable electromagnetic response (effectively, a phase shift).
It is assumed that the DMA meta-elements are attached in groups in $N_{\rm RF}$ RFCs, so that all $N_{E} \triangleq N/N_{\rm RF}$ elements of each DMA row are connected to a common RFC. By modeling the signal propagation inside the DMA microstrip (assuming negligible attenuation) and adopting a Lorentzian-constrained phase model for its unit elements, the frequency response $w_{i,j}$ of the element located in the $i$-th row ($i=1,\dots, N_{\rm RF})$ and $j$-th column ($j=1,\dots,N_{E})$ of the metasurface's grid can be compactly expressed as follows~\cite{DMA_Nir}:
\begin{align}\label{eq:dma-response}
    w_{i,j} \triangleq 0.5\left(\jmath + \exp{(\jmath \phi_{i,j})}\right) \exp{(\jmath \rho_{j} \beta)}, 
\end{align}
where $\phi_{i,j} \in [- \pi/2, \pi/2]$ denotes the \textit{controllable} phase shift induced by the DMA for the respected element, $\rho_{j}$ is the horizontal position of the $j$-th element in its microstrip, and $\beta$ is the wavenumber associated with the internal propagation, which is expressed as $\beta \triangleq (2 \pi / \lambda )\sqrt{\epsilon}$ with $\lambda$ being the wavelength of the carrier signal and $\epsilon$ denotes the permittivity of the microstrip (usually, $\epsilon=6$~\cite{DMA_Nir}).
For notation purposes, let us denote the concatenated vector of phase shifts as:
\begin{equation}
    \boldsymbol{\phi} \triangleq [ \phi_{1,1}, \phi_{1,2}, \dots, \phi_{2,1}, \dots, \phi_{N_{\rm RF}, N_{E}} ]^T,
\end{equation}
and the corresponding vector of the DMA configuration as
\begin{equation}
    \boldsymbol{w} \triangleq [ w_{1,1}, w_{1,2}, \dots, w_{2,1}, \dots, w_{N_{\rm RF}, N_{E}} ]^T.
\end{equation}
The phase shift configuration of the overall DMA satisfies the mean power constraint $\mathbb{E} \left[ \| \boldsymbol{w} \|^2 \right] \leq P_{\rm max}$. From now on in this paper, we will be referring to the elements of $\boldsymbol{\phi}$ and $\boldsymbol{w}$ as $\phi_n$ and $w_n$ ($n=1,\dots,N$), respectively, where $n = (i-1)N_{E}+j$.

At each localization request time slot $t$, the user is located at a 2D position $\boldsymbol{p}^t$ and transmits a fixed pilot symbol $x$ with unit power under a multipath channel. Orthogonal Frequency-Division Multiplexing (OFDM) with $L$ subcarriers is assumed and the channel at each $t$-th time slot and $l$-th subcarrier with $l=1,2,\ldots,L$ is modeled by the vector $\boldsymbol{h}^{t,l} \in \mathbb{C}^{N\times1}$. By using the notation $\boldsymbol{\tilde{w}}_i^t \triangleq \boldsymbol{w}^H_{(i-1)N_E+1:iN_E} \in \mathbb{C}^{1\times N_E}$ for the sub-vector of the configuration associated with the $i$-th DMA RFC at the $t$-th time slot, and representing the associated channel sub-vector as $\boldsymbol{\tilde{h}}_i^{t,l} \triangleq \boldsymbol{h}_{(i-1)N_E+1:iN_E}^{t,l} \in \mathbb{C}^{N_E\times1}$, the received signal at  the $i$-th RFC of the DMA at the access point at the $l$-th subcarrier and $t$-th time slot can be expressed as:
\begin{align}\label{eq:received-signal}
    y_{i}^{t,l} \triangleq \boldsymbol{\tilde{w}}_i^t \boldsymbol{\tilde{h}}_i^{t,l} x + \boldsymbol{\tilde{w}}_i^t \boldsymbol{\tilde{n}}^{t,l},  
\end{align}
where $\boldsymbol{\tilde{n}}^{t,l} \sim \mathcal{CN}(\mathbf{0}_{N_E},\sigma^2\mathbf{I}_{N_E})$ denotes the thermal noise vector at the receiver with $\mathbf{0}_{N_E}$ and $\mathbf{I}_{N_E}$ being the $N_E$-element zero vector and $N_E\times N_E$ identity matrix, respectively. Let us finally denote the concatenated receive vector from all $N_{\rm RF}$ DMA's RFCs as $\boldsymbol{y}^{t,l} \triangleq [y^{t,l}_1,y^{t,l}_2, \ldots, y^{t,l}_{N_{\rm RF}}]^T$.

\begin{figure}[t]
    \centering
    \label{fig:system-setup}
    \includegraphics[width=0.9\linewidth]{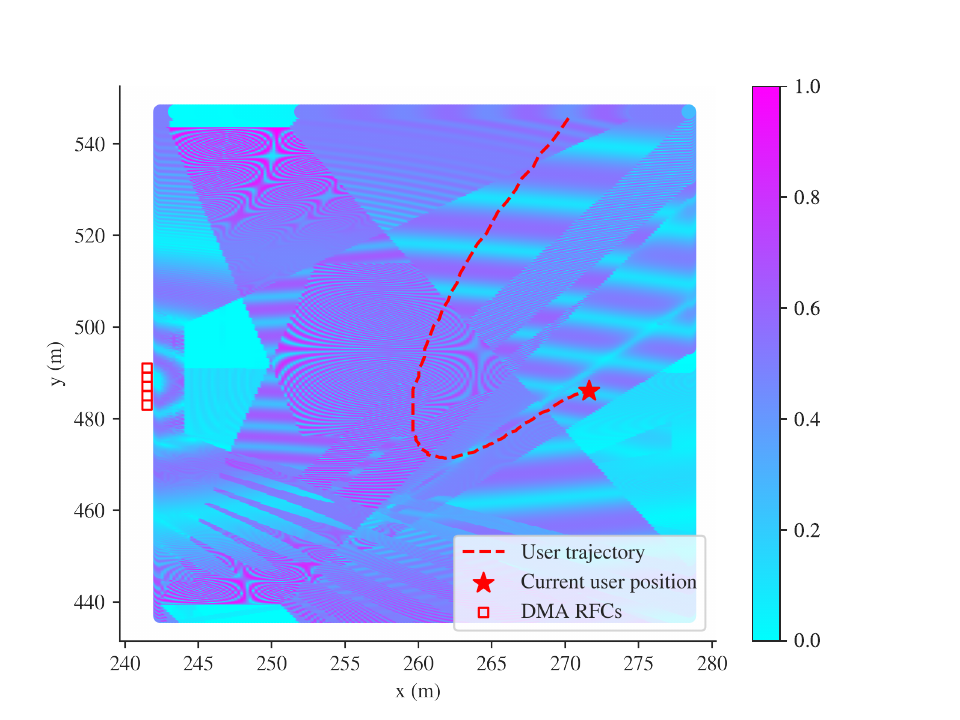}
    \vspace{-0.5cm}
    \caption{Top view of the non-LoS tracking scenario ``O1'' with one blocking and two main reflecting wall surfaces. The receiving DMA is positioned at $(235,485)$. The figure is colored by the received signal strength values, as provided via ray-tracing from \cite{DeepMIMO} using $3$ non-LOS channel paths. Values shown in normalized units. The red curve illustrates a randomly generated user trajectory of $100$ location points of $20$~cm separation.}
    
\end{figure}

To accurately model the non-LoS and multipath nature of the considered localization setting, we adopt the urban outdoor ``O1 - Blocked'' scenario of the DeepMIMO simulator~\cite{DeepMIMO} for the carrier frequency $f_c=28$~GHz and $500$~KHz bandwidth. This simulator provides ray-tracing-derived path details (see~\cite{WithRay} for a relevant survey with similar simulators) for user positions along a grid with $20$~cm separation. A top view of the simulation setup is given in Fig~\ref{fig:system-setup}.

\subsection{Tracking Objective}
Let us denote as $\mathcal{T}$ the trajectory instance within a predefined space containing the $|\mathcal{T}|$ sequential user positions. In this paper, we make use of the Bézier curves \cite{Graphics_Book} with $5$ control points for each trajectory by further discretizing each random user position to the closest one appearing on the DeepMIMO data set. The applied localization schemes output an estimated user position $\boldsymbol{\hat{p}}^t$ at every time slot $t$, therefore, the instantaneous localization error is given by $ e_{\rm pos}^t \triangleq \sqrt{ \|\boldsymbol{p}^t - \boldsymbol{\hat{p}}^t \|^2}$ and our proposed tracking objective entails minimizing the mean of this error within a trajectory. In mathematical terms, we consider the optimization problem:
\begin{align}\label{eq:tracking-problem}
    \mathcal{OP} : & \underset{\boldsymbol{ \{\hat{p}}^t\}_{t=1}^{|\mathcal{T}|}}{\min } ~ \frac{1}{|\mathcal{T}|} \sum_{t=1}^{|\mathcal{T}|} e_{\rm pos}^t.
\end{align}

\section{Proposed User Tracking Methodology}
In this section, a two-stage approach for online user tracking with a DMA-based receiver is presented. We first describe the channel estimation process, followed by the overall user tracking procedure. Then, the details of the proposed mMHSA NN that maps channel estimates with position estimations, and the considered AR model that leverages the latter estimations to better fit user mobility patterns, are provided.

\subsection{Channel Estimation}
By setting $\phi_{i,j} = - \pi/2 \triangleq \phi^0$ in~\eqref{eq:dma-response}, the response of each ($i,j$)-th DMA element gets zeroed. Using this setting for all indices $i$ and all $j \neq j'$, while simultaneously letting $\phi_{i,j'} \neq \phi^0$, the received signal in~\eqref{eq:received-signal} at each $i$-th DMA RFC at the $l$-th subcarrier and $t$-th time slot can be expressed as:
\begin{equation}
    y_{i}^{t,l} = \tilde{w}_{i,j'}^t \tilde{h}_{i,j'}^{t,l} x + \tilde{w}_{i,j'}^t \tilde{n}_{i,j'}^{t,l},
\end{equation}
which indicates that only the $j'$-the element of the $i$-th microstrip contributes to the received signal. By digitally removing the known effects of $\tilde{w}_{i,j'}^t$ and $x$ from $y_{i}^{t,l}$, an estimate for $\tilde{h}_{i,j'}^{t,l}$ can be obtained with an associated estimation error given by $\tilde{w}_{i,j'}^t x \tilde{n}_{i,j'}^{t,l}$.
By sequentially setting $j'=1,\dots,N_E$, an estimate of the complete channel vector is attained using $N_E$ repeated pilot signals.
We note that for practical purposes, $N_E$ is expected to be $o(N)$, thus, the overhead of this pilot-assisted channel estimation procedure is limited.

\subsection{Tracking Procedure}\label{sec:Tracking-procedure}
Assume a user moving along its predefined trajectory at some arbitrary time step $t$.
The first stage of the proposed user tracking approach involves training a regression NN, termed modified Multi-Head Self-Attention (mMHSA), that receives a channel measurement matrix as described above and outputs an initial position estimate $\boldsymbol{\tilde{p}}^t = f_{\boldsymbol{\theta}}^{\rm mMHSA}(\boldsymbol{H}^t)$,
which is parameterized by an arbitrary vector of weights, $\boldsymbol{\theta}$.
The mMHSA network disregards any temporal correlation appearing in tracking problems.
To that end, the second stage of the procedure involves the refinement of those initial estimates by accounting for the user's learned mobility patterns.
Concretely, assume that, at time $t$, mMHSA has provided estimates for all past and current positions $\{\boldsymbol{\tilde{p}}^{k}\}_{k=1}^{t}$.
An AR model is then applied on those initial predictions to obtain the final output position estimate $\boldsymbol{\hat{p}}^t$, as $\boldsymbol{\hat{p}}^t = f_{\gamma, \boldsymbol{z}}^{\rm AR}(\{\boldsymbol{\tilde{p}}^{k}\}_{k=1}^{t})$, so that correlations between past estimates can be learned.

Since this approach falls under supervised learning, both models require a training procedure with known user positions.
To that end, assume that $D$ trajectories of length $\mathcal{|T|}$ are collected as $\mathcal{D} = \{ \{ (\boldsymbol{H}^{t,d}, \boldsymbol{p}^{t,d})\}_{t=1}^{|\mathcal{T}|} \}_{d=1}^{D}$ and a shuffled version of $\mathcal{D}$ is compiled as $\mathcal{D'} = \{ (\boldsymbol{H}^{q}, \boldsymbol{p}^{q})\}_{q=1}^{|\mathcal{T}|D}$.
The mMHSA is, thus, trained via standard Stochastic Gradient Descent (SGD) \cite{Goodfellow-et-al-2016} over $\mathcal{D'}$ on the loss function $J_{\mathcal{D'}}(\boldsymbol{\theta}) = 1/({|\mathcal{T}|D})\sum_{q=1}^{|\mathcal{T}|D} \sqrt{\|\boldsymbol{p}^{q} -  f_{\boldsymbol{\theta}}^{\rm mMHSA}(\boldsymbol{H}^q)\|^2}$ which minimizes the tracking objective \eqref{eq:tracking-problem}.
Once mMHSA is trained, it is used to predict the (known) positions for all channel instances in $\mathcal{D}$, to obtain $\{ \{ \boldsymbol{\tilde{p}}^{t,d}\}_{t=1}^{|\mathcal{T}|} \}_{d=1}^{D}$.
Therefore, the AR model is trained via SGD to minimize the error in the estimation between the ground-truth positions and the initial estimates of mMHSA as $J_{\mathcal{D}}(\gamma, \boldsymbol{z}) = 1/({|\mathcal{T}|D})\sum_{d=1}^{D}\sum_{t=1}^{|\mathcal{T}|} \sqrt{\|\boldsymbol{p}^{t,d} -  f_{\gamma, \boldsymbol{z}}^{\rm AR}(\{\boldsymbol{\tilde{p}}^{k,d}\}_{k=1}^{t}) \|^2}$.

\subsection{Modified Multi-Head Self-Attention Network (mMHSA)}

\begin{figure}[t]
    \centering
    \includegraphics[width=0.9\linewidth]{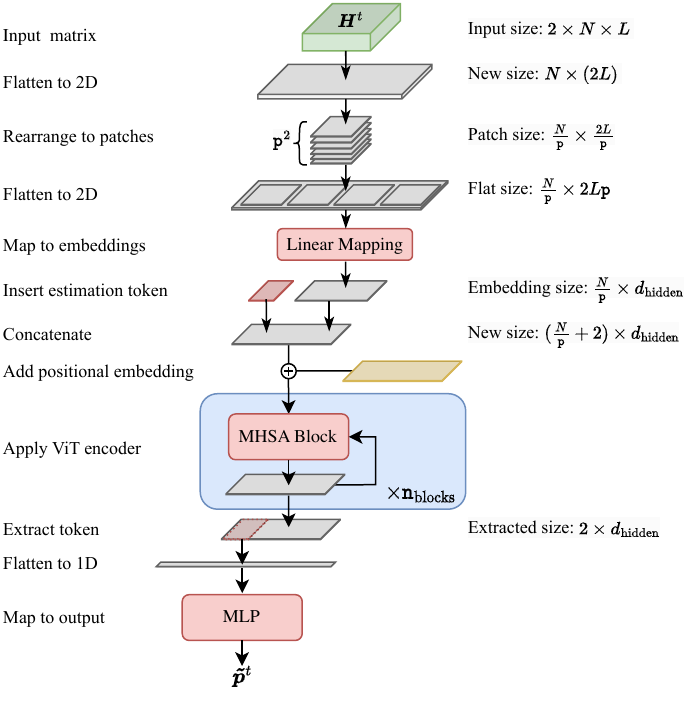}
    \vspace{-0.4cm}
    \caption{The proposed mMHSA architecture for position estimation based on noisy channel estimates, inspired by ViT \cite{ViT}. The design of the patches, estimation token, and positional embeddings have been modified from the original architecture. Red rectangles signify learnable blocks.}
    \label{fig:ViT-architecture}
\end{figure}

The mMHSA step aims to map noisy channel estimates to position estimations.
Accounting for the $N$ DMA meta-elements and the $L$ OFDM subcarriers, and by separating the real and imaginary channel components along a matrix dimension, the estimated channel at each $t$-th time slot can be expressed as a 3D matrix $\boldsymbol{H}^t \in \mathbb{R}^{2 \times N \times L}$, which resembles the matrix dimensions commonly appearing in computer vision applications.
It is noted that, due to the geometric nature of the considered channel model, the $\boldsymbol{H}^t$ elements are expected to exhibit high spatial correlation~\cite{correlated_Nakagami}.
Under the latter observations, our mMHSA architecture uses components of the state-of-the-art Vision Transformer (ViT), which was originally proposed for image classification tasks~\cite{ViT}.
The NN architecture is included in Fig.~\ref{fig:ViT-architecture} following the ViT naming and components.
In particular, each input $\boldsymbol{H}^t$ is first rearranged to two dimensions of size $(N, 2L)$, which is then sliced into $\mathtt{p} \times \mathtt{p}$ smaller patches.
The patches are then concatenated and passed through a linear mapping layer to attain the initial embeddings of dimension $d_{\rm hidden}$.
Two learnable tokens are appended in the embeddings instead of a single class token, since our architecture is intended for prediction of 2D vectors.
Positional embeddings are then added, similar to~\cite{ViT}, although the frequency of the embeddings is set to $2N$ instead of $10000$ in our architecture.
The embeddings are then forwarded to $\mathtt{n}_{\rm blocks}$ sequential transformer encoder blocks.
Each encoder is comprised an mMHSA layer followed by a Multi-Layer Perceptron (MLP), as proposed in~\cite{Attention}.
Finally, the 2D estimation tokens are extracted along each embedding dimension and an MLP head is responsible for extracting the position estimation $\boldsymbol{\tilde{p}}^t$ out of them.

\subsection{AR-Model-Based Position Estimation}
This stage receives the estimates $\boldsymbol{ \{\tilde{p}}^t\}_{t=1}^{|\mathcal{T}|}$ of the user position along a trajectory of $\mathcal{T}$ sequential user positions (i.e., $|\mathcal{T}|$ mMHSA outputs) and feeds an AR model with an exponentially decaying weighting factor to refine the position estimate at each $t$-th time slot as:
\begin{equation}\label{eq:ar-model}
    \boldsymbol{\hat{p}}^t = f_{\gamma, \boldsymbol{z}}^{\rm AR}(\{\boldsymbol{\tilde{p}}^{k}\}_{k=1}^{t}) =  \sum_{k=1}^{t} {\rm pow}({\rm \sigma}(\gamma), t-k+1) \boldsymbol{z} \circ \boldsymbol{\tilde{p}}^k,
\end{equation}
where $\gamma$ and $\boldsymbol{z} \in \mathbb{R}^2$ denote the model's parameters, which are optimized through SGD over a data set of collected user trajectories and corresponding mMHSA predictions as described in Section~\ref{sec:Tracking-procedure}.
In this expression, $\sigma(\cdot)$ denotes the sigmoid function that ensures that the decaying factor resides in $(0,1)$, $\circ$ indicates element-wise multiplication, and ${\rm pow}(\cdot,\cdot)$ denotes exponentiation.

\section{Numerical Evaluation}
For the numerical evaluation, we adopt DeepMIMO's ``O1 - Blocked'' scenario~\cite{DeepMIMO}, where the base station with index $3$ plays the role of the receiving DMA positioned at $(235,485,6)$.
The geometry is given in Fig.~\ref{fig:system-setup}, where approximately $10^6$ user positions are sampled at a spacing of $20$~cm inside a $110 \times 40$~m outdoor urban area.
The user travels along $75000$ randomly generated 2D trajectories of $|\mathcal{T}|=100$ positions at a mean separation of $2.5$ m, remaining at a fixed height of $2$~m.
We keep $N = 256$, $L=4$, and $N_{\rm RF}=16$ as fixed values and we set the estimation noise floor power per symbol to $-60$~dBm.
The mMHSA network is built with $3$ Transformer Encoder blocks with $2$ heads and a hidden embedding dimension of $16$.
A learning rate of $0.002$ and a training period of $60$ epochs were used for both models.
Only half the data set was used for training the mMHSA, and the rest was used for testing to limit the number of random trajectory positions that are seen by the network during training, regardless of the internal regularization components.

Since typical localization and tracking works assume LoS \cite{LIS_Positioning, ActiveSensingICC23, DA-MUSIC, CRB_DoA_Review} or more involved systems and procedures that don't facilitate real time user tracking \cite{CS_3D_Distributed, MetaSensingDRL}, we have implemented a fingerprinting localization scheme \cite{Fingerprinting} as a baseline for position estimation. Disregarding trajectories and estimation errors, the Received Signal Strength Indicators (RSSIs) for all $N_{\rm RF}$ RFCs and $L$ subcarriers corresponding to the data points of the training set are stored in a data base.
To make the position estimation, this baseline observes the current RSSI from the user's pilots and outputs the position of the closest RSSI match in the data base (i.e. nearest neighbor heuristic).

\begin{figure}[t]
    \centering
    \includegraphics[width=0.86\linewidth]{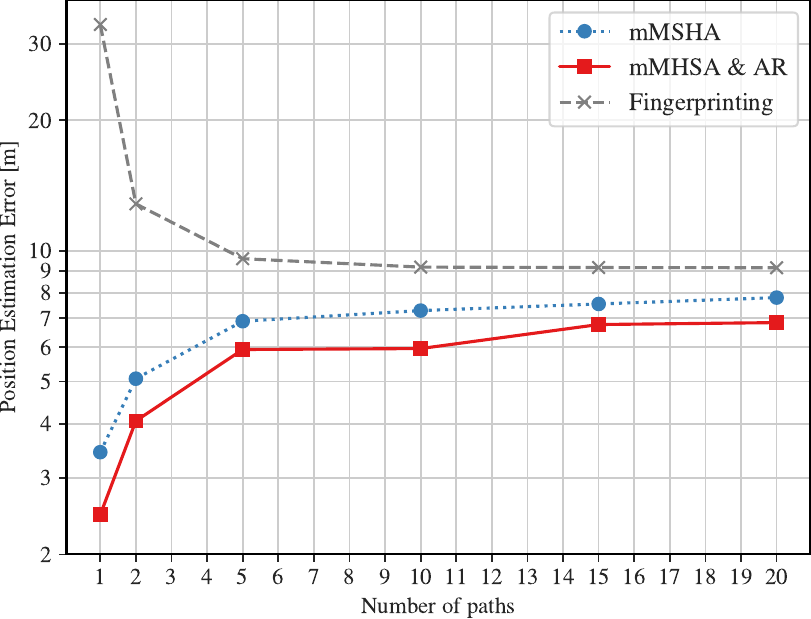}
    \vspace{-0.4cm}
    \caption{Localization error of the proposed methodology and baseline over different number of signal paths.}
    \label{fig:loc-vs-paths}
\end{figure}

The position estimation errors along $25000$ randomly sampled test trajectories are given in Fig.~\ref{fig:loc-vs-paths} over increasing numbers of signal paths, both for the initial mMHSA estimations, the two-step mMHSA and AR process, and the fingerprinting baseline.
It can be inferred that the proposed architecture has substantial overheads over the baseline, especially when fewer reflection paths are present.
The performance difference is decreasing as the effects of multipath get more pronounced, with the mMHSA-AR methodology resulting in smaller position errors nonetheless.
The addition of the AR step is beneficial to the overall scheme by reducing the estimation error roughly $1$-$2$~m in all cases.
When a single reflection path is considered, this approach is able to attain a mean position error of $2.47$~m.
Finally, it is worth stating that while the presence of multipath has an adverse effect in the estimation accuracy, the induced errors nearly plateau after a certain threshold of $5$ paths.

\vspace{-0.3cm}
\section{Conclusion}
\vspace{-0.2cm}
In this paper, the problem of online user tracking via a receiving DMA has been considered, dealing specifically with mobile users in non-LoS multipath environments.
The DMA architecture was used to provide noisy estimates of the channel with low pilot overheads, which are then fed into a variation of the ViT architecture to map frequency responses to user position estimates.
The tracking problem was handled by an AR model over initial estimates along sampled user trajectories to further reduce the estimation error based on mobility patterns.
The numerical evaluation has illustrated that this approach is able to attain high localization performance in a large area non-LoS scenario with multipath channels derived from the DeepMIMO ray-tracing simulator.
This approach was able to attain a mean position error below $2.5$~m when a single channel path is present, consistently outperforming the fingerprinting baseline in all cases, while the effects of the increasing number of paths are shown to be bounded.


\FloatBarrier
\bibliographystyle{IEEEtran}
\vspace{-0.1cm}
\bibliography{references}

\end{document}